\pdfoutput=1
\documentclass{article}
\newif\iffullversion
\fullversiontrue

\usepackage[preprint]{neurips_2024}

\usepackage[utf8]{inputenc} %
\usepackage[T1]{fontenc}    %
\usepackage{hyperref}       %
\usepackage{url}            %
\usepackage{booktabs}       %
\usepackage{amsfonts}       %
\usepackage{nicefrac}       %
\usepackage{microtype}      %
\usepackage{xcolor}         %

\usepackage{xurl}
\usepackage{comment}
\usepackage{graphicx} %
\usepackage{multirow}
\usepackage{xspace}
\usepackage{wrapfig}
\usepackage{booktabs} %
\newcommand{\namedref}[2]{\hyperref[#2]{#1~\ref*{#2}}\xspace}

\newcommand{\figureref}[1]{\namedref{Fig.}{fig:#1}}

\newcommand{\tableref}[1]{\namedref{Table}{tab:#1}}

\newcommand{\sectionref}[1]{\namedref{\S}{sec:#1}}
\newcommand{\appendixref}[1]{\namedref{Appendix}{app:#1}}

\newcommand{\toolname}{\textsc{Marill}\xspace}

\usepackage{enumitem}
\usepackage{xcolor}
\usepackage{amsmath, amssymb}
\usepackage{scalerel}
\usepackage{subcaption}
\usepackage{tikz}

\newcommand{\numlayers}{N}
\newcommand{\tuned}{f}
\newcommand{\seqlen}{b}
\newcommand{\heads}{h}

\newcommand{\headdim}{d}

\newcommand{\bbR}{\mathbb{R}}
\newcommand{\calL}{\mathcal{L}}
\newcommand{\attn}{\mathsf{attn}}
\newcommand{\logits}{\mathsf{logits}}
\newcommand{\mse}{\mathsf{MSE}}
\newcommand{\nll}{\mathsf{NLL}}

\newcommand{\merge}{m}
\newcommand{\party}{\mathcal{P}}

\newenvironment{boxfig}[2]{\begin{figure}[#1]\fbox{
    \begin{minipage}{0.95\linewidth}
    \vspace{0.2em}\makebox[0.025\linewidth]{}    \begin{minipage}{0.9\linewidth}{{\small #2 }}
    \end{minipage}\vspace{0.2em}\end{minipage}}}{\end{figure}}

\newenvironment{boxfigfull}[2]{\begin{figure*}[#1]\fbox{
    \begin{minipage}{0.95\linewidth}
    \vspace{0.2em}\makebox[0.025\linewidth]{}    \begin{minipage}{0.9\linewidth}{{\small #2 }}
    \end{minipage}\vspace{0.2em}\end{minipage}}}{\end{figure*}}

\newcommand{\pprotocol}[4]{
\begin{boxfig}{!ht}{
\begin{center}
\textbf{#1}
\end{center}
    #4
\vspace{0.2em} } \caption{\label{#3} #2}
\end{boxfig}
}

\newcommand{\protocol}[4]{
\pprotocol{#1}{#2}{#3}{#4} }

\newcommand{\functionality}[4]{
\pprotocol{#1}{#2}{#3}{#4} }

\title{MPC-Minimized Secure LLM Inference}

\author{%
Deevashwer Rathee$^{1}$\thanks{Equal contribution} \quad Dacheng Li$^{1\ast}$ \quad Ion Stoica$^1$ \quad Hao Zhang$^2$ \quad Raluca Ada Popa$^1$ \\
$^1$UC Berkeley \quad $^2$UC San Diego \\
\href{mailto:deevashwer@berkeley.edu}{\texttt{deevashwer@berkeley.edu}} \\
}

\begin{document}

\maketitle

\begin{abstract}
Many inference services based on large language models (LLMs) pose a privacy concern, either revealing user prompts to the service or the proprietary weights to the user.
Secure inference offers a solution to this problem through secure multi-party computation (MPC), however, it is still impractical for modern LLM workload due to the large overhead imposed by MPC.
To address this overhead, we propose \toolname, a framework that adapts LLM fine-tuning to minimize MPC usage during secure inference.
\toolname introduces high-level architectural changes during fine-tuning that significantly reduce the number of expensive operations needed within MPC during inference, by removing some and relocating others outside MPC without compromising security.
As a result, \toolname-generated models are more efficient across all secure inference protocols and our approach complements MPC-friendly approximations for such operations.
Compared to standard fine-tuning, \toolname results in $3.6-11.3\times$ better runtime and $2.4-6.9\times$ better communication during secure inference across various MPC settings, while typically preserving over $90$\% performance across downstream tasks.

\end{abstract}

\section{Introduction} \label{sec:intro}
Transformer-based large language models (LLMs) have revolutionized machine learning (ML).
Since the announcement of ChatGPT, we have seen the release of a plethora of proprietary LLMs (e.g., GPT-4~\cite{openai2023gpt4turbo}, Claude 2~\cite{anthrophic2024claude}, Bard~\cite{google2024bard}), as well as open-source LLMs (e.g., Llama~\cite{touvron2023llama}, Mistral~\cite{jiang2023mistral}) that are now competitive against their proprietary counterparts~\cite{chiang2024chatbot, wang2023decodingtrust, berkeley-function-calling-leaderboard, liu2024your}. %
Recently, companies have started to finetune these models on domain-specific data to improve their performance on downstream tasks such as chatbots, virtual assistants, and copilots~\cite{openai2023gpt4turbo, anyscale2024fine, cohere2024command}. 

Using these finetuned models to power such user-facing services, however, raises significant privacy concerns. On one hand, the providers of these finetuned models do not want to expose their models' weights, as these models are often trained on proprietary data and represent competitive differentiation. On the other hand, users do not want to send their queries to these providers as these queries might contain sensitive or proprietary information (e.g. IP-protected code or user data).
In fact, some enterprises prohibit their users from using LLM services, e.g., Samsung recently banned the use of external LLM services after an employee accidentally leaked sensitive code to ChatGPT~\cite{samsung-leak}.

Secure inference is a promising solution to address this challenge as it can provide privacy for both parties through secure multi-party computation (MPC)~\cite{gmw, yaogc}.
There is a long line of work on secure inference~\cite{secureml, delphi, cryptflow2, sirnn, securenn, cryptgpu, llama-secure, orca} offering different performance and security tradeoffs, with the recent work focusing on secure transformer inference~\cite{mpcformer, crypten-characteristic, puma, bumblebee, ciphergpt, sigma}.
In principle, the service provider can use any of these recent secure inference protocols to support its privacy-preserving service.
However, despite massive strides in efficiency, these protocols are still impractical for today's LLMs. For instance, the state-of-the-art solution~\cite{sigma} requires $23$ s and $15.9$ GB of communication for the first token generation on a small $137$M parameter model with $1024$ input tokens. We expect the runtime and communication to degrade to around $6.5$ minutes and $240$ GB for a more typical $7$B parameter model, which is impractical.

\begin{figure*}[t]
    \centering
    \includegraphics[width=0.72\textwidth]{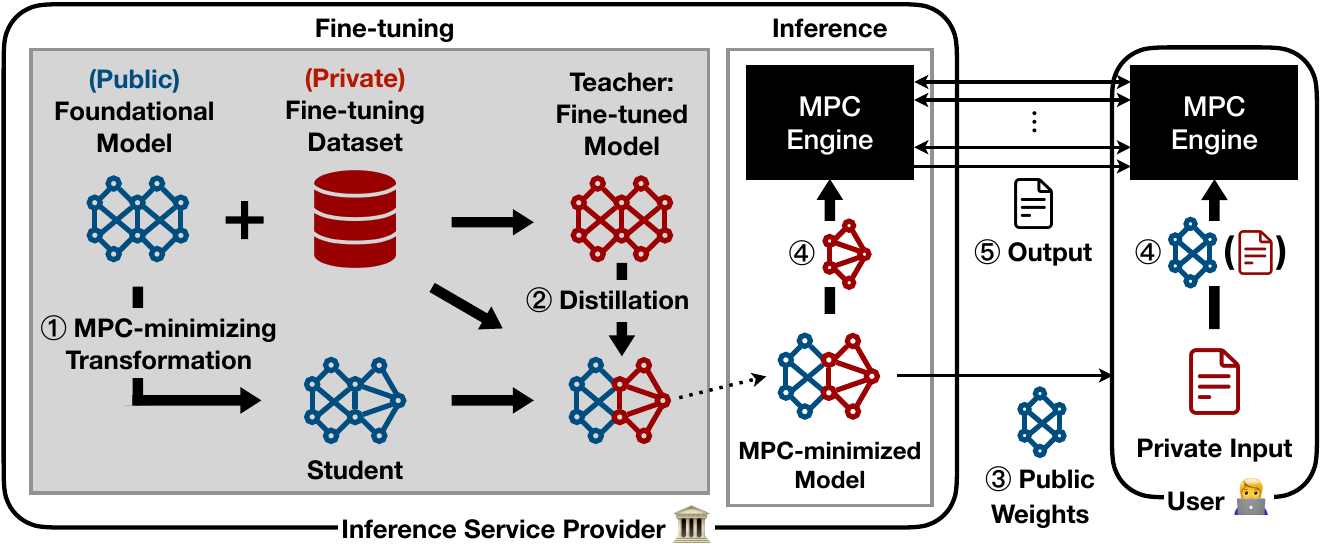}
    \caption{End-to-end workflow of our system. The private and public components are highlighted in red and blue, respectively. The gray region represents our fine-tuning framework, \toolname, that outputs an \emph{MPC-minimized} inference model. Note that \toolname differs from prior works such as MPCFormer~\cite{mpcformer} since they output a (fully) fine-tuned model after fine-tuning. Consequently, the inference phase (steps 3-5) in our system also differs from the prior works in two ways: (i) only a part of the inference model is private, and thus, only that part is fed to the MPC engine, and (ii) instead of directly feeding its private input, the client inputs the partial inference result of the model's public component on its private input.
    In the figure, we only show single token generation; subsequent tokens can be generated similarly since the client has access to all the tokens generated so far.
    Additionally, we only show two parties each running an MPC engine instance. Orthogonally, there is also an additional helper party in some protocols that helps speed up secure inference~(\appendixref{mpc-setting}).
    }
    \vspace{-2mm}
    \label{fig:overview}
\end{figure*}

To minimize this overhead, prior works have focused on low-level operations that are expensive to compute within MPC, and have proposed \emph{MPC-friendly} approximations for those operations~(\sectionref{related-work}).
In this work, we consider an orthogonal approach targeting high-level architectural changes, that offer a complementary way to minimize the MPC overhead.
Instead of simplifying operations, such architectural changes reduce the number of expensive low-level operations needed within MPC. Critically, this strategy does not (necessarily) eliminate these operations from the inference process entirely; rather, it relocates them outside of MPC without compromising security, where their cost is relatively negligible. 
Our work is the first to explore this high-level strategy, which we term \emph{MPC-minimization}.
We achieve this through fine-tuning, and our key insight is that \emph{fine-tuning, when carefully tailored to secure inference, can unlock significant opportunities for MPC-minimization}.

Following this insight, we propose a fine-tuning framework \toolname\footnote{\toolname stands for \underline{M}PC-Minimized \underline{AR}chitecture for Secure \underline{I}nference of \underline{LL}Ms} that makes strategic modifications to standard fine-tuning, guided by the unique characteristics of MPC.
The fine-tuned models output by \toolname are (i) MPC-minimized while maintaining the same level of security~(\sectionref{threat-model}), and (ii) achieve ML performance close to that of standard fine-tuned models through knowledge distillation~(\sectionref{techniques}).
Crucially, since \toolname essentially compresses the model within MPC, the resulting models are significantly more efficient across all secure inference protocols (\sectionref{secure-inference-perf}). Furthermore, as mentioned earlier, \toolname introduces only high-level architectural changes that complement MPC-friendly approximations. We demonstrate that integrating these approximations with \toolname leads to further efficiency improvements~(\sectionref{prior-approx}).
Now, we present a brief overview of our techniques and the model component (in bold) they minimize within MPC:
\begin{itemize}[leftmargin=*]
    \item \textbf{Leveraging open-sourced models}:
    As alluded to earlier, open-source LLMs have become more powerful and are now competitive against proprietary models~\cite{chiang2024chatbot, wang2023decodingtrust, berkeley-function-calling-leaderboard, liu2024your}. Consequently, a trend has emerged where an increasing number of service providers opt to fine-tune these open-source models with their private datasets instead of pre-training their own proprietary models~\cite{anyscale2024fine, cohere2024command}.
    Standard fine-tuning updates all the model weights with the private data, necessitating the entire model to run within MPC and precluding any potential benefits of the publicly available pre-trained weights.
    In light of this, we propose two fine-tuning strategies that effectively leverage the public weights to minimize MPC: 
    \begin{itemize}[leftmargin=*]
        \item \textbf{Layer Freezing}~(\sectionref{layer-freezing}):
        We reduce the \textbf{number of transformer layers} that need to be evaluated within MPC by restricting fine-tuning updates (and thus, private weights) to just the final layers of the pre-trained model.
        We resort to such strict demarcation because alternating private and public layers still require the bottleneck operations in the public layers to run within MPC~(\sectionref{inference-overhead}), and simply pruning the public layers leads to poor task performance~(\sectionref{ablation}).
        \item \textbf{Low-rank Adaptation (LoRA)}~(\sectionref{lora-adaptation}):
        Recent parameter-efficient fine-tuning techniques like LoRA~\cite{lora} have shown that it is possible to achieve comparable task performance by training only a small fraction of the model's weights.
        Although LoRA was designed to expedite the fine-tuning process, we demonstrate a novel application of LoRA and show that it can be repurposed to reduce the \textbf{dimensions of matrix multiplications} within MPC -- a runtime bottleneck in the natural two-party setting as well as during decoding~(\ref{app:inference-stages})-stages) in other MPC settings~(\sectionref{lora-adaptation}).
    \end{itemize} 
    \item \textbf{Reducing self-attention overhead}:
    We analyzed the cost profile of LLM inference under various MPC settings and found that the self-attention module is the bottleneck in the most efficient settings~(\sectionref{head-merging}). The standard solution to minimize self-attention operations is to employ head-pruning~\cite{head-pruning}. However, we have to prune up to $75\%$ heads (and their corresponding parameters) to achieve significant improvements and we find that this leads to a large accuracy drop despite fine-tuning~(\sectionref{ablation}). To address this loss, we introduce the following novel technique:
    \begin{itemize}[leftmargin=*]
        \item \textbf{Head-merging}~(\sectionref{head-merging}):
        We reduce the \textbf{number of attention heads} within MPC by merging $\merge$ heads into one, but simultaneously, we also increase the head dimension proportionally to preserve all the parameters.
        While it seems that we did not gain anything because the computational FLOPs remain the same, we show that head-merging actually matches the performance of head-pruning within MPC~(\sectionref{ablation}).
        This is based on the key observation that the self-attention operations that are the bottleneck in MPC only scale with number of heads and not the head dimension.
        Our experiments show that if the heads are merged carefully, head-merging achieves much better task performance than head-pruning~(\sectionref{ablation}).
    \end{itemize}
\end{itemize}

The end-to-end workflow of \toolname is summarized in \figureref{overview}.
Compared to standard fine-tuning, \toolname-generated models have $3.6-11.3\times$ faster runtime and $2.4-6.9\times$ lower communication across state-of-the-art secure inference frameworks in various MPC settings~ (\sectionref{secure-inference-perf}).
We evaluate the ML performance of \toolname on three different kinds of tasks, namely, code generation~\cite{chen2021evaluating}, chatbot~\cite{mt-bench}, and machine translation~\cite{kocmi2022findings}. Across these benchmarks, we show that \toolname typically preserves over $90\%$ of the standard fine-tuned performance~(\sectionref{ml-perf}).

\section{Related Work} \label{sec:related-work}
\textbf{Secure Inference Protocols.}
In this work, we focus on MPC-based secure inference protocols for neural networks which started with the seminal work of SecureML~\cite{secureml}.
SecureML considers the two-party setting that only involves the service provider and the client, and after many follow-up works in this setting~\cite{secureml, gazelle, minionn, delphi, cryptflow2, sirnn, gala, cheetah, heliks, iron, ciphergpt, bumblebee, bolt}, the performance has improved by orders of magnitude.
Despite these improvements, 2PC still poses very large overheads. Thus, subsequent works have considered other settings that introduce an additional helper party such as 3PC with honest majority~\cite{securenn, cryptflow, chameleon, aby3, falcon, puma} and 2PC with trusted dealer (2PC-Dealer)~\cite{crypten, llama-secure, orca, sigma}. 
Other works have accelerated secure inference protocols by leveraging GPU acceleration~\cite{crypten, cryptgpu, piranha, orca, sigma}. 

Recent work~\cite{iron, ciphergpt, bumblebee, bolt, puma, crypten-characteristic, sigma, roman2023fhe} in all these settings have focused on secure transformer inference since they represent the majority of the AI workload today.
Our work is orthogonal to these protocols and can be used to accelerate secure inference with any of them~(\appendixref{marill-scenario}). %

\textbf{MPC-friendly Approximations.}
Several works~\cite{mpcformer, secureml, cryptonets, cryptonas, faster-cryptonets, the-x, delphi, secformer, jha2021deepreduce, peng2023rrnet, cho2022selective, reagendeep, lou2020safenet, kundu2023learning, zhang2023sal} have proposed approximate implementations for non-linear activations like softmax and GeLU to make them more MPC-friendly.
These approximations typically introduce a large drop in model performance.
MPCFormer~\cite{mpcformer} proposed a two-stage distillation process to bridge this gap.
Majority of these works~\cite{delphi, jha2021deepreduce, ghodsi2020cryptonas, peng2023rrnet, cho2022selective, reagendeep, kundu2023learning, lou2020safenet, zhang2023sal} also use Neural Architecture Search (NAS) to employ multiple approximations within the same network depending on the precision level required.

Our work is complementary to these approximations as we make high-level changes to the architecture, as opposed to the underlying operations.
We show in \sectionref{prior-approx} that these approximations can be combined with \toolname to yield further performance improvements.
Additionally, \toolname differs from these works in two key aspects: (i)
while these works output models where all weights are private, \toolname produces models that have a mix of public and private weights,
and (ii) the model architecture in NAS-based works depends on the private training data and leaks additional information, whereas \toolname is statically configured independent of the training data. %

\section{Threat Model} \label{sec:threat-model}

We inherit the threat model from prior secure LLM inference works which all assume a semi-honest (or passive) adversary that follows the protocol exactly but tries to learn information about the private inputs from the messages it sees during the protocol.
This adversary controls an \emph{unknown} subset of the MPC participants, where the size of the subset is defined by the MPC setting.
Like prior works, we also assume that the model architecture is public and the service only wants to hide the model weights.
We formally prove security in \appendixref{security-proof}.
We note that our work is not limited to a semi-honest adversary and discuss extensions to malicious security in \appendixref{malicious-security}.

\section{Performance Characteristics of Secure Inference} \label{sec:inference-overhead}

Secure inference relies on secure multi-party computation (MPC)~\cite{gmw,yaogc}, a cryptographic primitive that allows mutually distrusting parties to compute any function on their private inputs without revealing anything beyond the function output.
Prior secure inference works, specifically, have considered three MPC settings~(\appendixref{mpc-setting}), each making different assumptions about the participants.
In this section, we highlight the unique cost profile of MPC in these settings and discuss how it motivates the design of our techniques in \sectionref{techniques}.

\textbf{Interaction costs.}
Unlike plaintext computation, most operations within MPC require interaction among the MPC participants. This imposes two additional performance overheads in addition to computation size, namely, \emph{communication size} and \emph{rounds of communication}.
For most MPC protocols, this cost of interaction ends up being the bottleneck and it is the primary reason why MPC is orders of magnitude slower than plaintext computation. %

\textbf{Multiplications with public weights come for free.}
Since MPC operates natively over integers, recent secure inference works use fixed-point representation to emulate real-number arithmetic. Additionally, prior works maintain the invariant that the intermediate state after every network layer is arithmetically secret-shared (ASS) among MPC participants. This approach minimizes the cost of arithmetic operations, such as integer multiplications and additions, which dominate ML workloads.
In an ASS scheme, a secret value $x$ is split among $n$ MPC participants such that (i) each party $\party_i$ receives a share $x_i$ and any set of $n-1$ shares reveals nothing about $x$, and (ii) the sum of all shares reconstructs the secret $x = x_1 + \ldots + x_n$. The linear nature of this reconstruction function allows secret-shared values to be added locally (without interaction) by simply adding the corresponding secret shares, making additions within MPC relatively so inexpensive that they are considered ``free".
Similarly, any affine operation with public coefficients on secret-shared values, such as a matrix multiplication with public weights, also becomes free. In \sectionref{lora-adaptation}, we show how low-rank adaptations can leverage this property to reduce the number of multiplications between secret-shared values. %

\textbf{Non-arithmetic operations are the bottleneck in the most efficient MPC settings.}
Non-arithmetic operations are used to implement comparisons in maxpool, activation functions such as ReLU and GeLU, exponentiation and division in softmax, as well as the truncation operations in fixed-point multiplications.
We analyzed state-of-the-art secure inference frameworks~(\sectionref{secure-inference-perf}) in the most efficient MPC settings, namely, 3PC and 2PC-Dealer~(\appendixref{mpc-setting}), and found that non-arithmetic operations account for over $88$\% of the runtime and communication during secure inference with a sequence length of 2048.
This is in stark contrast to plaintext computation where non-arithmetic operations have a minimal contribution to the total FLOPs and the inference latency.
Guided by this insight, we proposed head-merging in \sectionref{head-merging}, a technique that preserves the FLOPs and still yields significant performance improvements.

\textbf{A mix of public and private weights typically does not speedup secure inference.}
Since multiplications with public weights come for free, one would expect significant improvements to secure inference if most of the weights were public.
However, to preserve the standard guarantees of the MPC, an intermediate state that depends on both the private input and any private weight must not be revealed to any party.
Consequently, once the computation involves a single private weight, all subsequent non-arithmetic operations need to be performed within MPC, which as we just discussed are the bottleneck in the most efficient MPC settings for secure inference.
This restriction motivated the design of layer-freezing in \sectionref{layer-freezing}, which separates the public and private weights across layers such that the non-arithmetic operations in public layers are performed outside MPC.

\section{Techniques} \label{sec:techniques}
\begin{figure}[t]
    \centering
    \begin{subfigure}[b]{0.5\textwidth}
        \centering
        \includegraphics[width=0.7\textwidth]{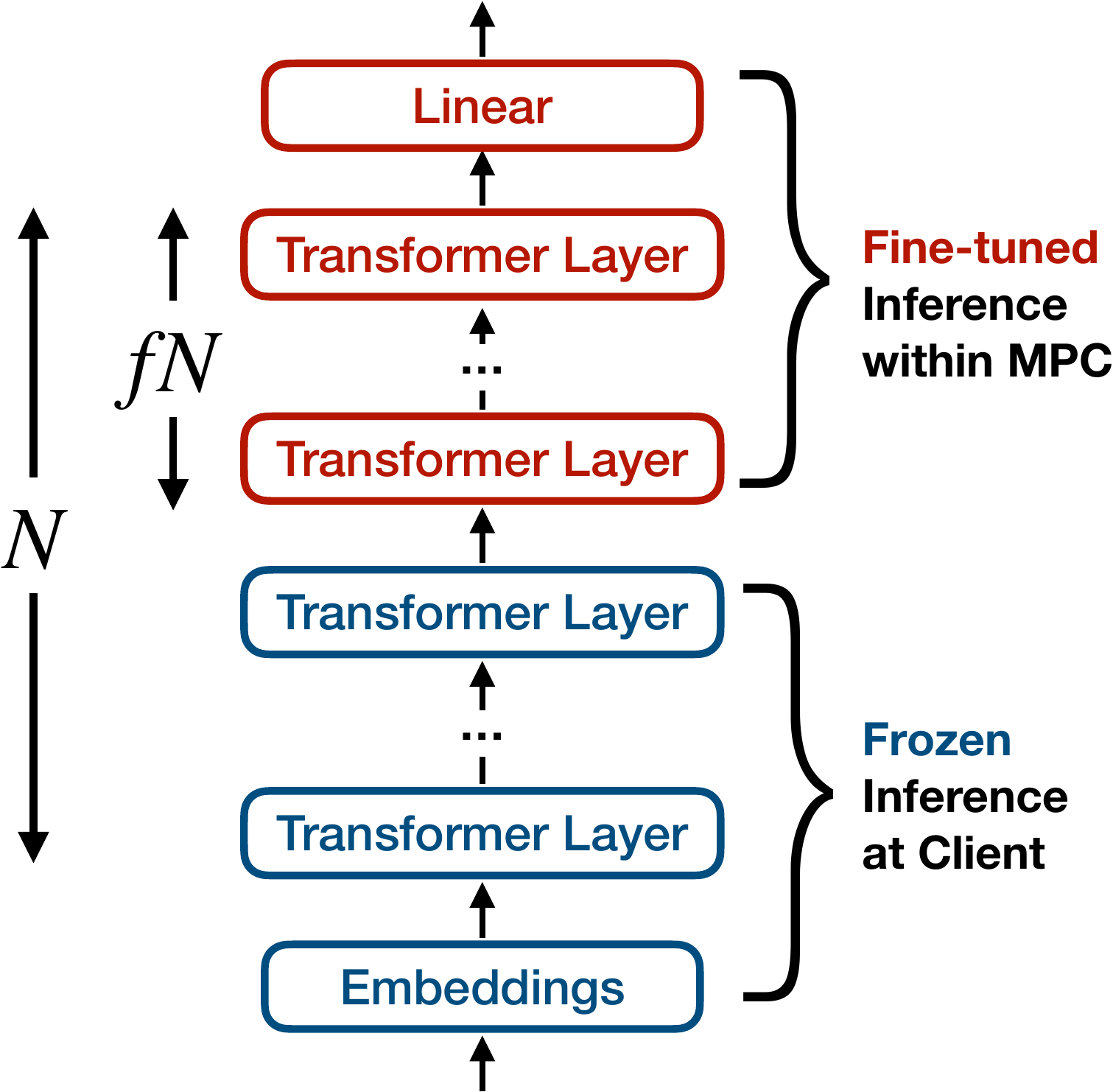}
        \caption{Layer Freezing with fraction $f$ layers fine-tuned} \label{fig:layer-freezing}
    \end{subfigure}
    \hfill
    \begin{subfigure}[b]{0.45\textwidth}
        \centering
        \includegraphics[width=0.7\textwidth]{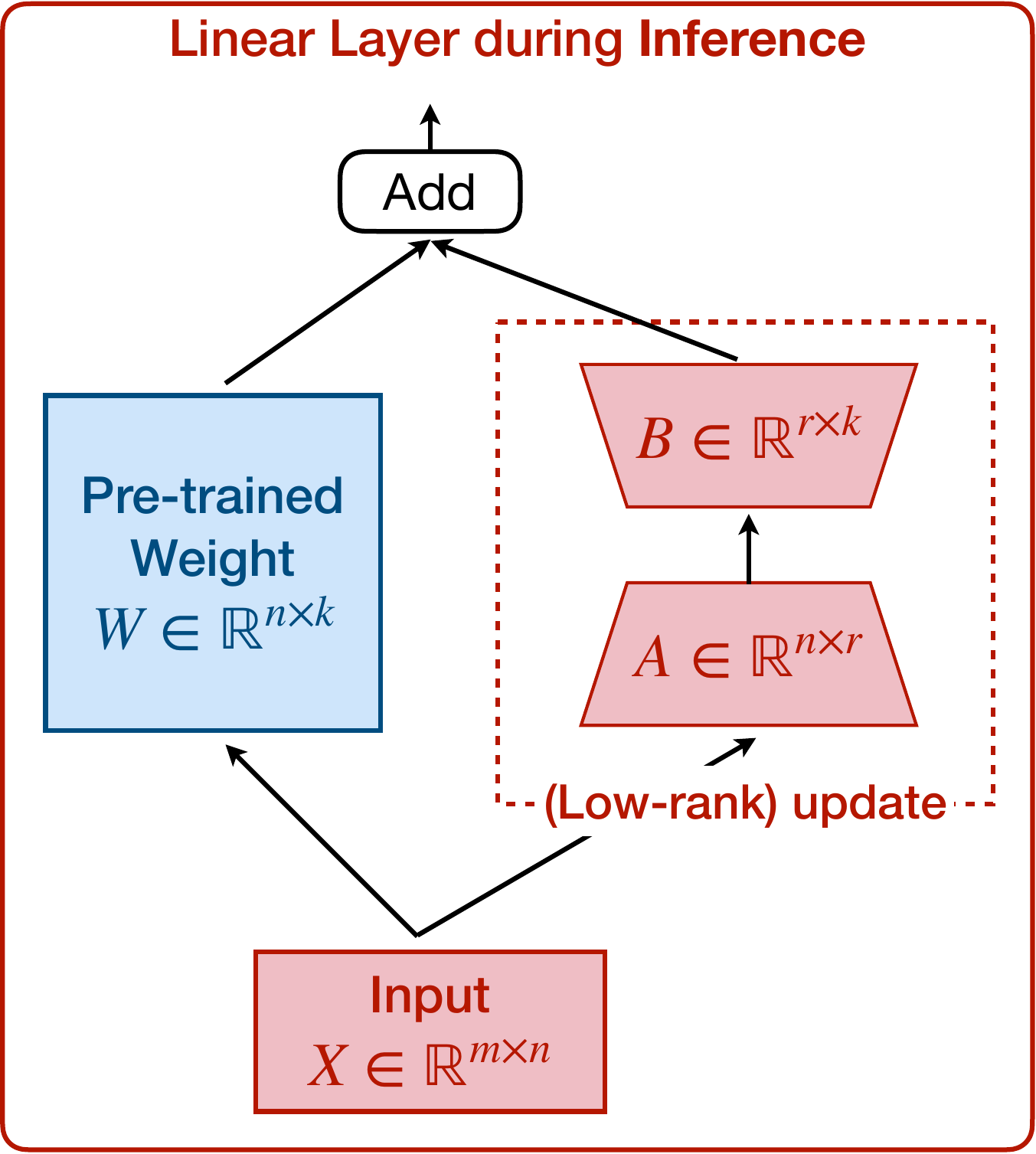}
        \caption{Linear layer during inference with LoRA} \label{fig:lora-adaptation}
    \end{subfigure}
    \caption{\toolname's techniques that leverage public weights (marked in blue).}
    \label{fig:open-source}
    \vspace{-2mm}
\end{figure}

In this section, we describe our techniques that minimize the need for expensive operations within MPC.
We start with layer-freezing (\sectionref{layer-freezing}) that reduces the number of layers evaluated within MPC. Next, we discuss LoRA~(\sectionref{lora-adaptation}) and head-merging (\sectionref{head-merging}) that minimize arithmetic and non-arithmetic operations, respectively, in the private layers.
Distillation details are deferred to \appendixref{distillation}.

\subsection{Layer Freezing} \label{sec:layer-freezing}
Our starting point is the observation that when an open-source model is fine-tuned on a private dataset, only the fine-tuned weights need to be kept private during inference.
To leverage this insight, consider using a technique from prior work that only fine-tunes a fraction of model weights~\citep{gandhi2023distil}.
However, as explained in \sectionref{inference-overhead}, these techniques typically do not significantly speed up inference.
This is because they update weights throughout the network, including near the input, which means that almost all non-arithmetic operations -- typically the bottleneck -- must be performed within MPC.

To this end, our solution (\figureref{layer-freezing}) effectively leverages public weights by deferring fine-tuning to only the final layers of the transformer, thereby also deferring MPC to these final layers. During inference, the client receives the weights for the bottom layers (identical to the open-source pre-trained model) from the server, computes the output of these layers locally, and then engages in MPC with the server for the top layers. Consequently, if only a fraction $f$ of the layers are fine-tuned, all MPC overheads are reduced by a factor of $\frac{1}{f} \times$~(\tableref{layer-freezing}).
Although delegating the computation of the bottom layers to the client might seem like a limitation, this approach actually \emph{reduces client overheads} by the same factor, since the MPC overhead on the client in secure inference protocols is orders of magnitude higher than the overhead of plaintext inference\footnote{The overhead on MPC participants, including the client, is nearly identical in all secure inference protocols, and even the most efficient secure inference protocol SIGMA has a $73\times$ overhead over plaintext inference~\cite{sigma}.}.

\subsection{LoRA Adaptation} \label{sec:lora-adaptation}
In \sectionref{inference-overhead}, we discussed how multiplication with public weights is free during secure inference. Here, we demonstrate how LoRA~\cite{lora}, a technique developed for parameter-efficient fine-tuning, can be repurposed to minimize integer multiplications during inference. These operations account for up to $95\%$ of the runtime in the state-of-the-art 2PC work Bumblebee~\cite{bumblebee}. Beyond the 2PC setting, we found that multiplications also dominate the decoding~(see~\appendixref{inference-stages}) runtime in 3PC and 2PC-Dealer settings, which are otherwise bottlenecked by non-arithmetic operations (\sectionref{inference-overhead}).
This occurs because the linear layers during decoding perform matrix-vector multiplications instead of matrix multiplications, making key matrix-multiplication optimizations from~\cite{secureml} no longer applicable.

A LoRA adapter on a weight matrix $W \in \bbR^{n \times k}$ is a product of two low-rank matrices $A \in \bbR^{n \times r}$ and $B \in \bbR^{r \times k}$, where $r \ll \min(n, k)$.
During fine-tuning, only the low-rank matrices are updated, and at inference time, $A \times B$ is merged into the pre-trained weight $W$ to minimize inference overhead. This approach updates all the model weights and we do not get any benefit from the public pre-trained weights.
In our solution, we crucially \emph{do not} merge the product $A \times B$ with the pre-trained model weights and keep the matrices separate as shown in \figureref{lora-adaptation}.
To see why this reduces multiplications, consider the evaluation of a LoRA-adapted linear layer:
for input $X \in \bbR^{m \times n}$, the evaluation function can be written as $X \times (W + A \times B)$.
Na\"ively, the complexity of this expression is $O(m n k)$.
However within MPC, the product $X \times W$ comes for free~(\sectionref{inference-overhead}).
To evaluate the remaining expression $X \times A \times B$, instead of computing $A \times B$ first, we can first evaluate $X \times A$ and then multiply it with $B$. This reduces the overall complexity to $O(mr (n + k))$; for $n = k = 3200$ and $r = 64$, this idea reduces the number of multiplications by $25 \times$.

\vspace{-2mm}
\subsection{Head Merging} \label{sec:head-merging}
\begin{wrapfigure}{r}{0.50\textwidth}
    \centering
    \vspace{-4\intextsep}
\includegraphics[width=0.42\textwidth]{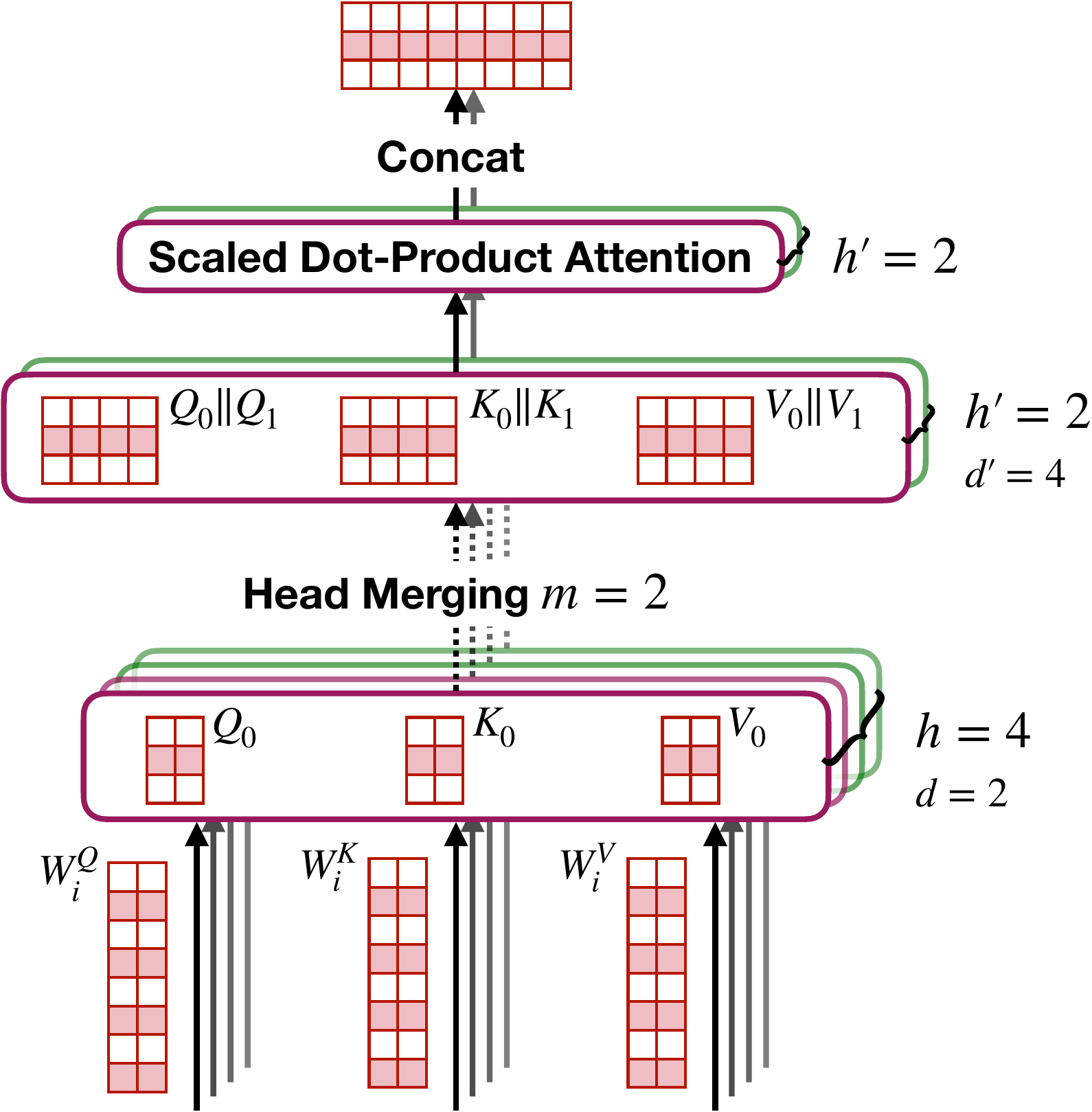}
    \caption{Head merging ($\merge = 2$) example for seq-len $\seqlen=3$, \#heads $\heads=4$, and head-dim $\headdim = 2$. After merging, $\heads$ reduces to $\heads' = 2$ and $\headdim$ increases to $\headdim' = 4$. The red matrices represent that head-merging is only performed in private layers.}
    \label{fig:head-merging}
    \vspace{-\intextsep}
\end{wrapfigure}
The most efficient secure inference works~\cite{puma, crypten, sigma} operate in the 3PC and the 2PC-Dealer settings~(\appendixref{mpc-setting}).
In these settings, non-arithmetic operations are the bottleneck.
Among these operations, those in the self-attention module are of particular interest because:
(i) the self-attention mechanism is the only component that scales quadratically with the sequence length $\seqlen$, (ii) the state-of-the-art works in both 3PC~\cite{puma} and the 2PC-Dealer~\cite{sigma} settings exhibit a super linear blowup in runtime when $\seqlen \geq 1024$, highlighting that self-attention is indeed the bottleneck for large $\seqlen$, and (iii) applications such as chatbots and copilots which have real-time requirements require a large sequence length. %
Thus, we focus on minimizing the non-arithmetic operations in the self-attention module in this work.

\textbf{Reducing number of heads.}
The self-attention mechanism has two non-arithmetic operations: (i) softmax, and (ii) truncations (from fixed-point multiplications), and the complexity for both is $O(\seqlen^2 \heads)$, where $\heads$ is the \#heads.
Hence, we seek to reduce $\heads$ by a factor $\merge$ so that all operations are reduced proportionally.
The standard technique for minimizing heads is head-pruning~\cite{head-pruning}, which analyzes the importance of each head over the training dataset, and prunes the insignificant heads.
This achieves our goal, however, we have to prune $75\%$ of the heads (as well as their parameters) for $\merge=4$, and this results in a large accuracy loss~(\sectionref{ablation}).

\textbf{Preserving the pre-trained parameters.}
To this end, we observe that unlike plaintext inference, FLOPs do not dictate the secure inference cost~(\sectionref{inference-overhead}) and it is possible to achieve similar speedups as head-pruning despite preserving all the parameters~(\sectionref{ablation}).
This is also evident in the complexity of non-arithmetic operations in self-attention, which are independent of the head-dimension $\headdim$.
Thus, we propose a technique called head-merging that reduces the number of heads $\heads$ by $\merge\times$, while simultaneously increasing the head dimension $\headdim$ proportionally, thereby preserving all parameters from the pre-trained model.
Specifically, $\heads$ heads are divided into groups of $\merge$, and the QKV matrices for heads within the same group are concatenated as shown in \figureref{head-merging}.
Concretely, given matrices $\{ Q_i, K_i, V_i \}_{i \in [\heads]}$ of dimension $\mathbb{R}^{\seqlen \times \headdim}$, the head attention outputs $\{ \mathsf{head}_j \}_{j \in [\heads/\merge]}$ after merging are as follows: %
$\mathsf{head}_j = \mathsf{softmax}\Big(\frac{\sum_{\ell=j\merge}^{(j+1)\merge} Q_\ell K^T_\ell}{\sqrt{\merge \headdim}}\Big) \cdot (V_{j\merge} \| \cdots \| V_{(j+1)\merge}) \in \mathbb{R}^{\seqlen \times \merge\headdim}$.

\textbf{Merging similar heads.}
In the expression above, adjacent heads are grouped such that heads $j\merge$ to $(j+1)\merge$ belong to group $j$. This strategy does not consider the similarity among heads, resulting in minimal accuracy improvement over head-pruning~(\sectionref{ablation}).
To group heads based on similarity, we follow the strategy from \cite{bian2021attention} that computes the pairwise Jensen-Shannon distance between all heads within the same layer.
Once we have the pairwise distances, we perform K-Medoid clustering~\cite{kaufman1990partitioning} to organize heads into $\heads/\merge$ groups.
Finally, to get groups of the same size, we redistribute heads based on a linear sum assignment that minimizes the sum of distances from the medoid within each group.
We found that merging similar heads using this method performs significantly better, leading to up to $8\%$ gain in accuracy~\sectionref{ablation}.

\section{Evaluation} \label{sec:evaluation}
\begin{figure*}[t]
     \centering
     \begin{subfigure}[b]{0.8\linewidth}
         \centering
         \includegraphics[width=\linewidth]{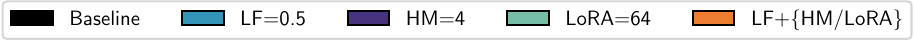}
         \vspace{-10pt}
         \label{fig:legend}
     \end{subfigure}
     \begin{subfigure}[b]{0.49\linewidth}
         \centering
         \includegraphics[width=\linewidth]{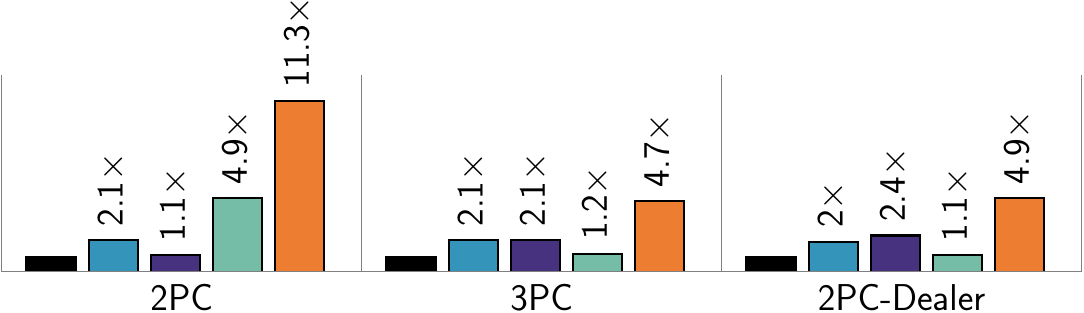}
         \caption{Pre-filling Time}
         \label{fig:prefilling-time}
     \end{subfigure}
     \hfill
     \begin{subfigure}[b]{0.49\linewidth}
         \centering
         \includegraphics[width=\linewidth]{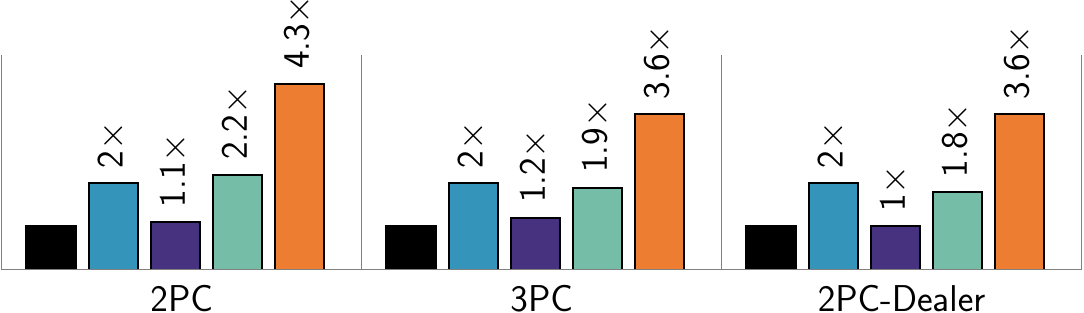}
         \caption{Decoding Time}
         \label{fig:decoding-time}
     \end{subfigure}
     \begin{subfigure}[b]{0.49\linewidth}
         \centering
         \includegraphics[width=\linewidth]{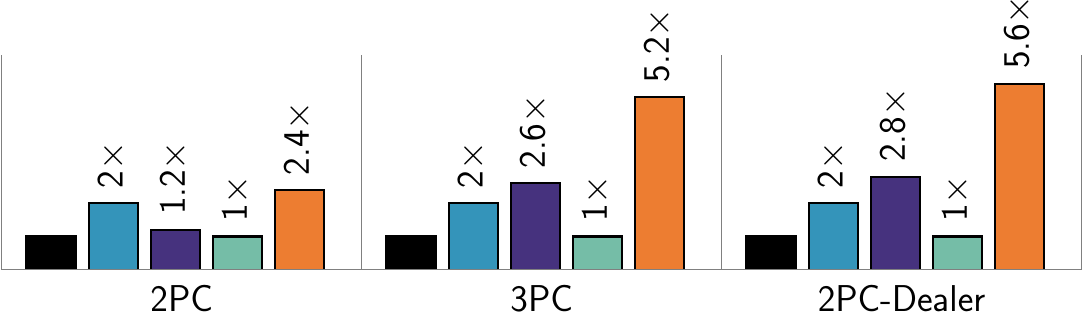}
         \caption{Pre-filling Communication}
         \label{fig:prefilling-comm}
     \end{subfigure}
     \hfill
     \begin{subfigure}[b]{0.49\linewidth}
         \centering
         \includegraphics[width=\linewidth]{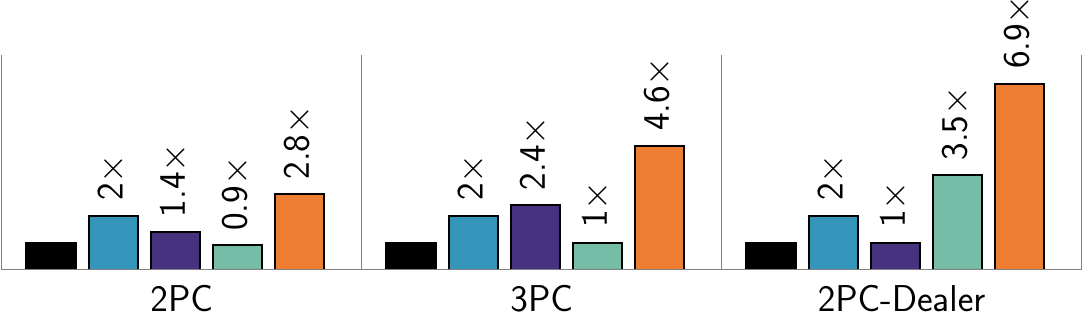}
         \caption{Decoding Communication}
         \label{fig:decoding-comm}
     \end{subfigure}
    \caption{Secure inference performance of \toolname vs standard fine-tuning for \texttt{openllama-3b-v2}.
    The sequence length is set to $b=64$ for 2PC and $b=2048$ for 3PC and 2PC-Dealer. 
    The numbers on the bars represent the improvement factor over the baseline.
    The final bar in each plot represents the combination of layer-freezing with head-merging or LoRA, whichever performs better independently.
    }
    \label{fig:mpc-exp}
    \vspace{-2mm}
\end{figure*}
In this section, we first evaluate the secure inference cost~(\sectionref{secure-inference-perf}) of \toolname-generated models and their ability to preserve ML performance~(\sectionref{ml-perf}).
Next, we perform the same analysis for prior MPC-friendly approximations integrated with \toolname~(\sectionref{prior-approx}).
Finally, we do an ablation study in \sectionref{ablation} that considers alternative designs for \toolname's techniques.  

\textbf{Secure Inference Setup.} %
We perform the secure inference experiments on state-of-the-art (open-sourced) frameworks in all MPC settings considered by prior work, namely, 2PC~\cite{bumblebee,spu}, 3PC~\cite{puma,spu}, and 2PC-Dealer~\cite{crypten,crypten-characteristic}.
The experiments were run on two or three machines (depending on the MPC setting) connected via LAN connection with $16$ Gbps bandwidth and $0.1$ ms latency.
Each machine was equipped with an Intel Xeon Platinum 8173M Processor with $16$ vCPUs, $128$ GB RAM, and a $V100$ GPU with $16$ GB memory.
Since the 2PC-Dealer framework~\cite{crypten} supports GPU acceleration, we ran it on the V100. Experiments on other MPC frameworks were run on CPU.
All experiments were multi-threaded.
All reported numbers consider end-to-end costs.

\textbf{Models and Datasets.}
We consider three privacy-sensitive tasks for LLMs: chatbot, coding, and machine translation.
For the chatbot task, we fine-tune \texttt{open-llama3b-v2} on the ShareGPT dataset and evaluate it on the MTBench dataset, following~\citep{mt-bench, li2023long}. OpenLLaMA is a popular open-source model that replicates the LLaMA model~\citep{openlm2023openllama, touvron2023llama}.
For the coding task, we fine-tune \texttt{deepseek-coder-1.3b-base} on the MagiCoder dataset~\cite{wei2023magicoder} and evaluate it on the HumanEval benchmark~\cite{chen2021evaluating}. For the machine translation task, we fine-tune \texttt{open-llama3b-v2} on the ParroT dataset~\cite{jiao2023parrot} and evaluate it on the WMT22 (De$\Rightarrow$En) benchmark~\cite{kocmi2022findings}.

\textbf{Fine-Tuning Hyperparameters.} We set the fine-tuning hyperparameters according to the papers that curated the corresponding fine-tuning dataset: \citep{mt-bench} for MTBench, \citep{wei2023magicoder} for HumanEval, and \citep{jiao2023parrot} for WMT22. We only vary the batch size and number of training epochs to better suit some techniques. For instance, we observed that LoRA favors a smaller batch size in our setting. We include the detailed hyperparameters in \appendixref{hyperparameters}.

\subsection{Secure Inference Performance}
\label{sec:secure-inference-perf}

In this section, we compare the secure inference performance of \toolname-generated models vs the baseline -- a fully fine-tuned model. \figureref{mpc-exp} summarizes these results for \texttt{openllama-3b-v2} as the pre-trained model. We first analyze the improvements from head-merging~(\sectionref{head-merging}) and LoRA~(\sectionref{lora-adaptation}) in the three MPC settings from prior work, and then discuss layer-freezing~(\sectionref{layer-freezing}) improvements.

\textbf{2PC}: LoRA improves the pre-filling runtime by $4.9\times$~(\figureref{prefilling-time}) because $92$\% of the 2PC runtime is spent in performing multiplications for \texttt{openllama-3b-v2} inference.
Decoding runtime is improved by $2.2\times$, which is less pronounced because the 2PC framework~\cite{bumblebee} does not amortize well over the smaller decoding computation.
In terms of communication, non-arithmetic operations are the bottleneck in 2PC, accounting for $72.5$\% of the total communication.
Still, we don't see a large improvement with head merging~(Figures~\ref{fig:prefilling-comm} \& \ref{fig:decoding-comm}) because it is designed for large sequence lengths and we could only run 2PC on small sequence lengths ($64$) due to its large memory requirements.

\textbf{3PC and 2PC-Dealer}:
Since non-arithmetic operations in the self-attention module become the bottleneck in these settings at large sequence lengths~(\sectionref{head-merging}), head-merging leads to runtime and communication improvements of $2.1-2.4\times$~(\figureref{prefilling-time}) and $2.6-2.8\times$~(\figureref{prefilling-comm}), respectively, in the pre-filling stage.
During decoding, integer multiplications are the runtime bottleneck instead~(\sectionref{lora-adaptation}), and hence, LoRA helps in this stage and we get $1.8-1.9\times$~(\figureref{decoding-time}) decoding runtime improvement.
In terms of decoding communication~(\figureref{decoding-comm}), 3PC exhibits a similar improvement as in pre-filling.
The communication improvement from LoRA for 2PC-Dealer is an implementation artefact\footnote{We had to employ matrix decomposition on all linear layers in the 2PC-Dealer setting to fit secure inference of (fully) fine-tuned LLaMA-3B on the V100 GPU.}.

\textbf{Layer Freezing (\sectionref{layer-freezing})}: We fine-tune half of the $26$ transformer layers in \texttt{openllama-3b-v2}. As expected, this leads to around $2\times$ improvement across the board: different settings, metrics, inference stages, and in combination with both techniques.
In some cases, layer freezing leads to a greater than $2\times$ improvement.
This is due to the omission of the embedding layer within MPC in addition to half of the transformer layers.
In general, we show in \tableref{layer-freezing} that layer freezing leads to $\frac{1}{\tuned}\times$ improvement in all metrics for a wide range of $\tuned$ values.
Overall, \toolname leads to $3.6-11.3\times$ better runtime and $2.4-6.9\times$ better communication across all MPC settings and inference stages.

\begin{figure}[t]
    \centering
\includegraphics[width=0.93\textwidth]{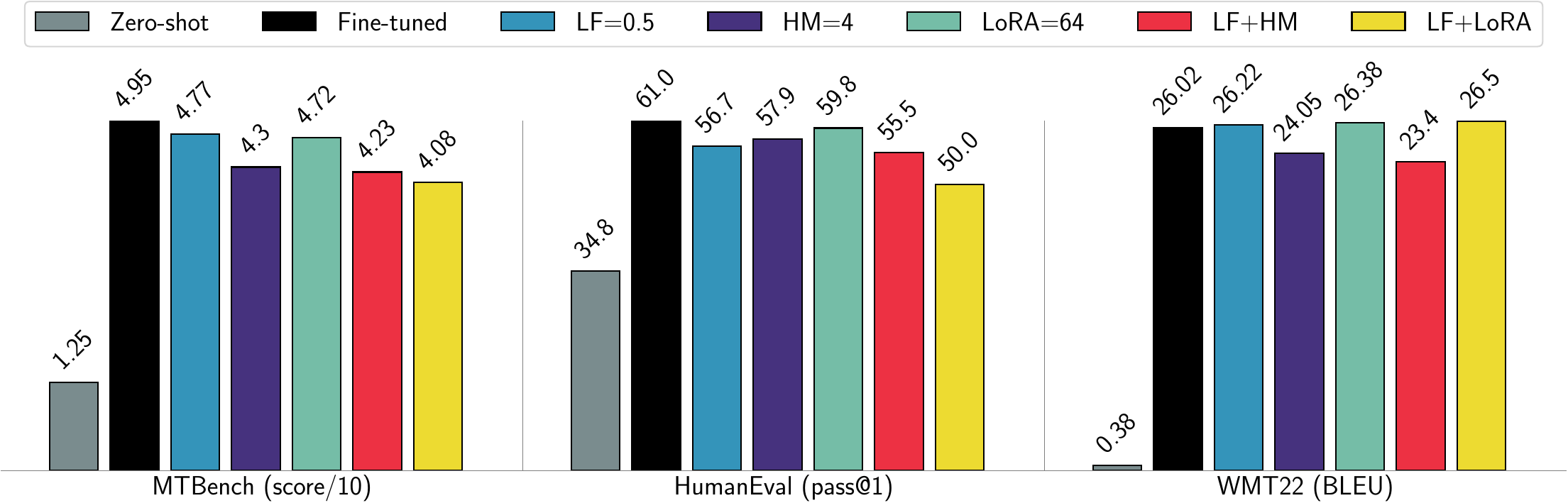}
    \caption{\toolname vs (fully) fine-tuned and zero-shot baselines.}
    \vspace{-4mm}
    \label{fig:ml_perf}
\end{figure}

\subsection{ML Performance} \label{sec:ml-perf}

\figureref{ml_perf} summarizes the ML performance of \toolname, the pre-trained model and the fully fine-tuned model on our three benchmarks.
First, we note that full fine-tuning significantly improves the performance of the pre-trained model across all three tasks.
\toolname's layer-freezing is also effective on all three tasks, preserving $93-100\%$ of the full fine-tuning performance.
On WMT and HumanEval benchmark, head-merging preserves $92-95\%$ performance, while on MTBench, it achieves $87\%$ performance.
The combination of layer-freezing and head-merging works well, incurring an additional loss of at most $4\%$ compared to head-merging alone. %
LoRA preserves over $95\%$ performance on all benchmarks.
While combining LoRA with layer freezing sometimes leads to a big drop in performance (MTBench and HumanEval), we note that using LoRA alone provides significant speed-ups, ranging from $2.2\times$ to $4.9\times$.
Overall, we observe that \toolname's techniques typically preserve over $90\%$ of the fully fine-tuned performance.

\begin{table}[t]
    \caption{HumanEval pass@1 performance of various techniques. The speedups and communication improvements are averages taken over the prefilling-stage in the 3PC and 2PC-Dealer settings.}
    \begin{subtable}{0.50\linewidth}
    \centering
    \caption{2ReLU approximation for softmax combined with \toolname(LF=0.5, HM=4)}
    \label{tab:marill-prior-approx}
    \footnotesize
    \begin{tabular}{lrrr}
    \toprule
    & & \multicolumn{2}{c}{Improvement} \\
               & pass@1 & Time & Comm.  \\
    \midrule 
    HM=4 & $57.9$ & $2.25\times$ & $2.7\times$ \\
    2ReLU + HM & $54.9$ & $3.25\times$ & $4.25\times$ \\
    \midrule
    LF=0.5 + HM=4 & $55.5$ & $4.8\times$ & $5.4\times$ \\
    2ReLU + LF + HM & $56.7$ & $6.9\times$ & $8.5\times$ \\
    \bottomrule
    \end{tabular}
    \end{subtable}
    \hspace{0.03\linewidth}
    \begin{subtable}{0.47\linewidth}
    \centering
    \caption{Adjacent/similar head-merging vs head-pruning (HP). Parameter denotes the head reduction factor.}
    \footnotesize
    \begin{tabular}{lrrr}
    \toprule
    & & \multicolumn{2}{c}{Improvement} \\
               & pass@1 & Time & Comm.  \\
    \midrule 
    HP=$4$ & $49.4$ & $2.45\times$ & $2.75\times$ \\
    HP=$2$ & $56.7$ & $1.7\times$ & $1.8\times$ \\
    \midrule
    HM=$4$ (adj.) & $50.0$ & $2.25\times$ & $2.7\times$ \\
    HM=$4$ (sim.) & $57.9$ & $2.25\times$ & $2.7\times$ \\
    HM=$2$ (sim.) & $60.4$ & $1.55\times$ & $1.8\times$ \\
    \bottomrule
    \end{tabular}
    \label{tab:head-pruning}
    \end{subtable}
    \vspace{-3mm}
\end{table}

\subsection{Integration of prior MPC-friendly approximations with \toolname} \label{sec:prior-approx}
In this section, we analyze the performance of \toolname when combined with prior MPC-friendly approximations, namely, Quad~\cite{mpcformer} and ReLU~\cite{chen2022x, zeng2023mpcvit} as GeLU/SiLU approximations, and 2Quad~\cite{mpcformer}, L2Quad~\cite{zhang2023sal} and 2ReLU~\cite{secureml} as softmax approximation.
First, we analyzed the ML performance of each approximation independently and found that the quadratic approximations from recent works led to a catastrophic loss on our benchmarks.
Specifically, on the HumanEval benchmark, Quad only achieves $31.7\%$ accuracy compared to $61\%$ of the baseline, and the fine-tuning diverges for L2Quad and 2Quad, resulting in $0\%$ accuracy.
In contrast, ReLU-based approximations work very well, with ReLU achieving the same accuracy as the baseline, and 2ReLU achieving $59.8\%$ accuracy.
Out of the ReLU-based approximations, only 2ReLU leads to significant efficiency improvements, with ReLU only improving the secure inference cost by at most $10\%$.
Thus, we only evaluate the combination of 2ReLU approximation with \toolname.

\tableref{marill-prior-approx} summarizes the accuracy results on the HumanEval benchmark and the corresponding secure inference improvements.
For the latter results, we focus on the 3PC and 2PC-Dealer settings because all prior approximations target non-arithmetic operations that are the bottleneck in these settings.
Our experiments show that 2ReLU works well with \toolname, incurring at most $3\%$ further accuracy loss on top of \toolname.
In exchange, 2ReLU improves \toolname's time and communication by $1.4-1.6\times$.
For reference, 2ReLU independently results in $1.95-2.15\times$ improvement over the baseline.
Overall, we get $6.9-8.5\times$ improvement in runtime and communication compared to the baseline, while still preserving over $90$\% of the baseline ML performance.

\subsection{Ablation Study} \label{sec:ablation}
\textbf{Layer-freezing vs layer-pruning.}
In layer-freezing, we froze the bottom layers of the transformer to move some layers outside of MPC. An alternative strategy to minimize layers within MPC is to simply prune some layers.
We experimented with layer-pruning on the HumanEval benchmark and evaluated the best-performing strategy from \cite{poor-mans-bert}, namely, top-layer pruning. 
For half of the layers pruned, we found that the accuracy drops from $61\%$ for the baseline to just $49.4$\% post layer-pruning.
In contrast, layer-freezing achieved an accuracy of $56.7\%$, a $12\%$ increase in relative performance, highlighting the importance of preserving the pre-trained model weights of the pruned layers.

\textbf{Head-merging vs head-pruning.}
We compared head-pruning~\cite{head-pruning} and head-merging~\sectionref{head-merging} on HumanEval, configuring head-pruning to prune the same number of heads from each layer so that it does not leak additional information about the private dataset.
\tableref{head-pruning} summarizes the results for both techniques when the heads are reduced by $2\times$ and $4\times$.
First, we note that head-merging achieves similar efficiency improvements to head-pruning for both head reduction factors, with head-pruning being at most $10\%$ faster and $2\%$ more communication efficient.
ML performance of head-merging, on the other hand, is much better since it preserves all the head parameters.
In particular, head-merging has up to $8\%$ better accuracy than head-pruning, and HM$=4$ even outperforms HP=$2$ in both ML and secure inference performance.
Note that these improvements only apply to similar head-merging, not adjacent head-merging, which na\"ively combines adjacent heads.
These results demonstrate the significance of preserving head parameters as well as merging heads based on similarity.

\section{Conclusion}
In this work, we designed a framework \toolname, that leverages open-sourced LLMs and introduces high-level architectural changes during fine-tuning to minimize MPC usage during secure inference.
We demonstrated that \toolname is effective in minimizing secure inference costs across MPC settings in exchange for a reasonable accuracy tradeoff. In particular, \toolname-generated models are $2.4 - 11.3\times$ more efficient for secure inference compared to a standard fine-tuned model, and they typically preserve over $90\%$ relative performance across multiple challenging LLM tasks.

\bibliography{references}
\bibliographystyle{plain}

\newpage
\appendix
\section{MPC Settings} \label{app:mpc-setting}
\begin{itemize}[leftmargin=*]
    \item \textbf{2-party computation (2PC)}: this setting assumes two MPC participants who do not trust each other, and thus, it is the most natural setting for secure inference.
    \item \textbf{Honest-majority 3-party computation (3PC)}: this setting has an additional helper party that also participates in MPC, and the adversary can corrupt at most any one of the three parties. Prior works considered this setting because having this helper party improves the MPC performance by orders of magnitude.
    \item \textbf{2PC with trusted dealer (2PC-Dealer)}: in this setting, there is an additional trusted dealer that is only responsible for distributing \emph{input-independent} correlated randomness to the computing parties in a pre-processing phase. The parties can then use this randomness to accelerate 2PC on their private inputs.
\end{itemize}

\section{LLM Inference Stages - Prefilling and Decoding} \label{app:inference-stages}

In this section, we briefly describe the two stages in LLM inference. Firstly, users provide a prompt in natural language to the system. The system then uses tokenizers to map the natural language into a vector $x_1, ... x_n$ %
through a process called tokenization~\cite{sennrich2015neural}.
Then the system performs the main inference process using LLMs. The inference process consists of two phases - the pre-filling phase and the decoding phase. Formally, the pre-filling phase computes probablity of the first token conditioned on the previous $n$ tokens $P(x_{n+1}|x_1, ... x_n)$~\cite{sheng2023fairness}. It then samples from the distribution and predicts the first token $x_{n+1}$. The decoding phase iteratively computes the next token based on the same logic. For instance, the first step in the decoding computes $P(x_{n+2}|x_1, ... x_{n+1})$ and samples to obtain $x_{n+2}$. The decoding phase terminate when the new token is an ending token, often referred to as the ``end-of-sentence" token (EOS). Interestingly, the left-to-right decoding nature has made the computation characteristics different~\citep{kwon2023efficient,yu2022orca,sheng2023fairness} in these two stages. Thus, we distinguish between the two phases when evaluating our techniques in this work. %

\section{Security Proof} \label{app:security-proof}
\newcommand{\pub}{\mathsf{pb}}
\newcommand{\prefilling}{\mathsf{prefill}}
\newcommand{\decoding}{\mathsf{decode}}
\newcommand{\update}{\mathsf{update}}
\newcommand{\evaluate}{\mathsf{evaluate}}
\newcommand{\pri}{\mathsf{pr}}
\newcommand{\func}{\mathcal{F}}
\newcommand{\client}{\mathcal{C}}
\newcommand{\server}{\mathcal{S}}
\newcommand{\arch}{M}
\newcommand{\ptarch}{M^\ast}
\newcommand{\param}{W}
\newcommand{\state}{\mathsf{st}}
\newcommand{\mpcparam}{\widetilde{W}}
\newcommand{\ptparam}{W^\ast}
\newcommand{\mpcarch}{\widetilde{M}}
\newcommand{\simulator}{\mathsf{Sim}}
\newcommand{\adversary}{\mathcal{A}}
\functionality{Secure Inference Ideal Functionality $\func_{\arch,n}$}{Ideal functionality for secure inference}{fig:secure-inference-func}{
This functionality is parameterized by the model architecture $\arch$ and \#outputs tokens $n$.
\begin{itemize}[leftmargin=*]
    \item \textbf{Client Prompt}: Receive prompt $p$ for $\arch$ from client $\client$, and store $p$ internally.
    \item \textbf{Server Weights}: Receive model weights $\param$ for $\arch$ from server $\server$, store $\param$ internally.
    \item \textbf{Pre-filling}: Perform pre-filling on the prompt to get state $\state \gets \arch.\prefilling(\param, p)$. Set $i \gets 0$.
    \item \textbf{Decoding}:
    If $0 < i < n$, receive token $x$ from the $\client$, update the state $\state \gets \arch.\update(\state, x)$, and increment $i$. Then, perform a decoding step on $\state$ to get an output token $y \gets \arch.\decoding(\state)$ and send $y$ to the client $\client$.
\end{itemize}
}

\protocol{\toolname's Secure Inference Protocol in the $\func$-hybrid model}{Our secure inference protocol.}{fig:secure-inference-prot}{

Let $\arch$ denote the entire model architecture (including LoRA and head-merging changes), $\arch_\pub$ denote the part of the architecture with public layers, and $\arch_\pri$ denote the part with private layers. Note that $\arch = \arch_\pub \| \arch_\pri$ due to the design of layer-freezing. Let $\param_\pub$ and $\param_\pri$ denote the corresponding weights for these parts. Client $\client$ has prompt $p$ and server $\server$ has weights $\param_\pri$. Both parties have $\param_\pub$. Let $n$ be the number of tokens to be generated.

\begin{enumerate}[leftmargin=*]
    \item Both parties initialize an instance of $\func_{\arch_\pri, n}$ and the $\server$ sends $\param_\pri$ to $\func_{\arch_\pri, n}$.
    \item The client locally evaluates the public part of the model on its prompt to get the hidden state for the prompt $h \gets \arch_\pub.\evaluate(\param_\pub, p)$, and sends $h$ to $\func_{\arch_\pri, n}$. Note that this is the input that $\arch_\pri$ expects to perform pre-filling on the prompt.
    \item $\client$ receives $y_1$ from $\func_{\arch_\pri, n}$.
    \item For $i = 2, \ldots, n$:
    \begin{enumerate}
        \item $\client$ locally evaluates the public part of the model on its prompt to get $h \gets \arch_\pub.\evaluate(\param_\pub, y_{i-1})$, and sends $h$ to $\func_{\arch_\pri, n}$. Note that this is the input $\arch_\pri$ expects to update its context state with $y_{i-1}$.
        \item $\client$ receives $y_i$ from $\func_{\arch_\pri, n}$.
    \end{enumerate}
    \item $\client$ outputs $(y_1, \ldots, y_n)$.
\end{enumerate}
}

\protocol{Simulator for \toolname's Secure Inference Protocol}{Simulator for \toolname's secure inference protocol.}{fig:secure-inference-sim}{
The simulator $\simulator$ internally runs the adversary $\adversary$, has access to its input prompt $p$ (since $\adversary$ is semi-honest), interacts with ideal functionality $\func_{\arch, n}$ on behalf of the party controlled by the adversary, and simulates $\func_{\arch_\pri, n}$ in the ideal-world. \\

\underline{If client $\client$ is corrupted}:
\begin{enumerate}[leftmargin=*]
    \item $\simulator$ sends prompt $p$ to $\func_{\arch, n}$ and receives $y_1$ from it.
    \item As $\func_{\arch_\pri, n}$, $\simulator$ receives $h$ from $\adversary$, ignores it, and sends $y_1$ to $\adversary$ as the output.
    \item For $i = 2, \ldots, n$:
    \begin{enumerate}[leftmargin=*]
        \item $\simulator$ sends $y_{i-1}$ to $\func_{\arch, n}$ and receives $y_i$ from it.
        \item As $\func_{\arch_\pri, n}$, $\simulator$ receives $h$ from $\adversary$, ignores it, and sends $y_i$ to $\adversary$ as the output.
    \end{enumerate}
\end{enumerate}

\vspace{5pt}
\underline{If server $\server$ is corrupted}:
\begin{enumerate}[leftmargin=*]
    \item Receive model weights $\param_\pri$ from $\adversary$, append it to the public weights $\param_{\pub}$ to get $\param = \param_\pub \| \param_\pri$ and forward $\param$ to $\func_{\arch, n}$. There is nothing else to simulate since the server does not receive any messages in our protocol in the $\func$-hybrid model.
\end{enumerate}
}

We prove the security of our protocol in the standard simulation paradigm~\cite{canetti00,gmw,lindellsim} that argues that whatever an adversary can learn in the real-world while interacting with honest parties in a protocol, it can also learn in an ideal-world while interacting with an ideal functionality that is incorruptible.
In particular, the proof shows that there exists a simulator in the ideal-world that can simulate the adversary's real-world view by only interacting with the adversary and the ideal-functionality.
Since the ideal functionality is designed to be trivially secure and not reveal anything about honest parties inputs beyond the function output, this proves that the adversary also can not learn this information from the actual protocol.
We describe the ideal functionality $\func$ that captures the security guarantees provided by any secure (transformer) inference protocol in \figureref{secure-inference-func}.
Note that the functionality does not leak any information to the server, and the client learns nothing beyond the output tokens.
The ideal functionality also allows the client to choose the latest token, which is not a problem in the semi-honest setting as the client will follow the protocol. We discuss how to ensure a malicious client inputs the right tokens in \appendixref{malicious-security}.
We designed \toolname to make black-box use of secure inference, and as such, we describe \toolname's secure inference protocol in the $\func$-hybrid model~\cite{canetti00,lindellsim} in \figureref{secure-inference-prot}.
Note that our protocol only focuses on layer-freezing because the other two techniques do not move operations outside of MPC, and thus, their privacy follows from the standard security of MPC.

Let $\arch$ be the model architecture (including LoRA and head-merging changes) and $\arch_\pri$ be the part of the architecture with private layers.
We prove that our protocol in the $\func_{\arch_\pri, n}$-hybrid model securely realizes the $\func_{\arch, n}$ functionality in the presence of a semi-honest adversary, where $n$ is the number of output tokens.
The proof is trivial and we describe the simulator for the sake of completeness in \figureref{secure-inference-sim}.

\section{Malicious Security} \label{app:malicious-security}
Our work is not limited to a semi-honest adversary and can also support a malicious adversary that deviates from the protocol arbitrarily.
Given a maliciously-secure protocol, our work inherits malicious security against the server directly as the server does not have any additional capabilities in our system. The simulator for a corrupted server also remains the same. 
Security against client needs careful assessment because the client in our system inputs a hidden state (output of a transformer layer), as opposed to a sequence of tokens in traditional secure LLM inference.
This does not impact semi-honest security because the client will follow the protocol and input the right hidden state.
However, a malicious client can input a state that doesn't correspond to any sequence of input tokens\footnote{The possible input token combinations are exponentially larger than the possible hidden states, even concretely at sequence lengths as small as $b=6$, but we do not know if transformer layers represent an onto function.} to potentially learn the model weights, or input a different token from what was generated to deviate the generation process.
This issue can be fixed by making the following changes to the protocol:
\begin{itemize}[leftmargin=*]
    \item In step 2, the client additionally provides a zero-knowledge proof-of-knowledge (ZKPoK)~\cite{zkp} proving that the hidden state it is secret-sharing corresponds to an actual sequence of tokens of the appropriate length.
    \item The secure inference protocol will output the token as well as a hiding commitment and its randomness to the client. Now, when the client will secret-share the hidden state for the latest token $y_{i-1}$ in step 4a, it'll additionally provide a ZKPoK proving that this state is consistent with the commitment received during the previous token generation.
    \item If either proof fails, the protocol will be aborted.
\end{itemize}

To complete the argument for malicious security, the ideal functionality and the simulator will be updated as follows:
\begin{itemize}[leftmargin=*]
    \item The ideal functionality $\func_{\arch, n}$ will track the generated tokens and abort if the token provided by the client $\client$ does not match the last generated token. In case $\arch$'s output and input don't match (as is the case for $\arch_\pri$), the functionality will be parameterized by the function $\arch_\pub$, and the functionality will check that $h$ received is equal to $\arch_\pub.\evaluate(y)$, where $y$ was the last generated token.
    \item Since the adversary is now malicious, the simulator does not have direct access to its input. Instead, the simulator will receive ZKPoK proofs in addition to the hidden states from the adversary $\adversary$. It will extract the adversary's input from these proofs. The rest of the simulation follows exactly the same way.
\end{itemize}

\section{Distillation} \label{app:distillation}
The modifications we make to the model for MPC-minimization change its learned knowledge during pre-training, and simply fine-tuning it leads to a large accuracy loss.
To bridge this accuracy gap, we turn to knowledge distillation (KD)~\cite{hinton-distill} in this work.

\figureref{overview} summarizes our distillation workflow.
First, we take the pre-trained model and apply the transformations that lead to an MPC-minimized architecture; the model thus obtained is the \emph{student}.
Then, we take the pre-trained model and fully fine-tune it to get the \emph{teacher model}, representing the performance baseline we want to match.
Finally, we use KD to ease the fine-tuning of the student model by matching its intermediate states with the teacher model.
The student model thus obtained can then be used for secure inference.

For layer-freezing and LoRA, we have a one-shot distillation procedure because they preserve the pre-trained knowledge.
Head-merging, on the other hand, requires a two-stage distillation process, similar in spirit to the strategy from MPCFormer~\citep{tinybert, mpcformer}.
Now, we describe the two stages of distillation. The configurations without head-merging only perform the second stage.

\begin{enumerate}[leftmargin=*]
    \item \textbf{Stage I - Attention and Hidden States KD}: to accommodate head-merging, we match the student and teacher outputs of MHA in each (trainable or private) transformer layer using the following loss function:
       $\calL_{\attn} = \sum_{i=\tuned\numlayers}^{\numlayers} \mse(\mathbf{a}_i^S, \mathbf{a}_i^T)$,
    where $\mathbf{a}_i^S$ and $\mathbf{a}_i^T$ are the MHA outputs in the $i$-th transformer layer of the student and teacher, respectively, $\tuned$ is the fraction of layers fine-tuned during training, and $\numlayers$ is the number of transformer layers. Similarly, we compute the distillation loss over hidden states after every (private) transformer layer: $\calL_{\mathsf{hidden}} = \sum_{i=\tuned\numlayers}^{\numlayers} \mse(\mathbf{h}_i^S, \mathbf{h}_i^T)$, where $\mathbf{h}_i^S$ and $\mathbf{h}_i^T$ are the hidden layer outputs in the $i$-th transformer layer of the student and teacher, respectively. For all experiments, we adopt coefficients $ \alpha_{\mathsf{attn}}$ and $ \alpha_{\mathsf{hidden}}$ for these two losses, and set them to $ \alpha_{\mathsf{attn}} = 0.1, \alpha_{\mathsf{hidden}} = 5.0$. We choose this value so that the two losses have similar magnitude, and we empirically observe that this brings the best accuracy. We skip this stage in experiments that do not use head-merging.
    \item \textbf{Stage II - Logits KD}: following stage I distillation, we employ supervised KD~\cite{hinton-distill, distil-bert} to match the student's token-level probability distribution (or logits) with that of the teacher. We use forward KL divergence (KLD) to measure the similarity of the distributions~\cite{gkd}. In addition to the distillation loss, we also minimize the negative log-likelihood (NLL) of the student's output on labels from the fine-tuning dataset. Overall, we use the following loss function in this stage:
       $\calL_{\logits} = \alpha_{\mathsf{KLD}} \cdot \mathsf{KLD}(\mathbf{z}^S, \mathbf{z}^T) + \alpha_{\mathsf{NLL}} \cdot \nll(\mathbf{z}^S, y)$,
    where $\mathbf{z}^S$ and $\mathbf{z}^T$ are the logits of the student and the teacher model, resp., $y$ is the label from the fine-tuning dataset, and $\alpha_{\mathsf{KLD}}$ and $\alpha_{\mathsf{NLL}}$ are scalar weights for the KLD and NLL terms, respectively. For all experiments, we set $\alpha_{\mathsf{KLD}} = 0.5, \alpha_{\mathsf{NLL}} = 0.5$.  %
\end{enumerate}

\textbf{Combining head-merging with other techniques.}
When using head-merging independently, we initialize the student weights with that of the teacher, perform a head similarity analysis on the teacher, and then perform the two-stages of distillation.
When head-merging is combined with layer-freezing, we perform the same procedure, except we replace teacher weights with the weights of the layer-freezing fine-tuned student.

\textbf{Other experiments.} Head-pruning and MPC-friendly approximations follow the same recipe as head-merging and require two-stage distillation. When combining MPC-friendly approximations with head-merging, we introduce them at the same time before stage I distillation.

\section{\toolname Configuration per Secure Inference Protocol} \label{app:marill-scenario}
\toolname's techniques target various potential bottlenecks that occur in secure inference protocols. In this section, we discuss which combination of techniques is the most suitable for a given secure inference protocol.
\begin{itemize}[leftmargin=*]
    \item If the protocol is bottlenecked on arithmetic operations, one should use LoRA because it provides an asymptotic reduction in these operations\footnote{Integer additions are also arithmetic but they have relatively negligible cost and can thus be ignored, leaving integer multiplications as the only arithmetic operation.}. %
    \item If the protocol is bottlenecked by non-arithmetic operations, consider the sequence length of the inference task. If the sequence length is large, prefilling will dominate the overall cost and self-attention will be the bottleneck. Head-merging will reduce all the non-arithmetic operations in self-attention and provide significant runtime and communication improvements. If the sequence length is small, decoding is likely to dominate the cost, and LoRA will present better runtime improvements.
    \item If there is no specific bottleneck, use layer-freezing and it will reduce overheads irrespective of the cost profile of the underlying protocol. For half the layer frozen, layer-freezing alone offers $2\times$ improvements across all inference scenarios and protocols. Otherwise, first apply one of the other two techniques, and then add layer-freezing on top.
\end{itemize}

\begin{table}[t]
\centering
\caption{Secure inference performance vs fraction of layers fine-tuned  $\tuned$.} \label{tab:layer-freezing}
\begin{tabular}{|c|c|c|c|c|c|c|}
\hline
Setting & $f=26/26$ & $f=13/26$ & $f=9/26$ & $f=6/26$ & $f=5/26$ & $f=4/26$ \\
\hline
\multicolumn{7}{|c|}{Prefilling Time} \\ \hline
2PC        & $1.0\times$ & $2.1\times$ & $2.9\times$ & $4.3\times$ & $5.1\times$ & $6.3\times$ \\ \hline
3PC        & $1.1\times$ & $2.1\times$ & $3.1\times$ & $4.6\times$ & $5.5\times$ & $6.9\times$ \\ \hline
2PC-Dealer & $1.0\times$ & $2.0\times$ & $2.9\times$ & $4.3\times$ & $5.1\times$ & $6.4\times$ \\ \hline
\multicolumn{7}{|c|}{Prefilling Comm} \\ \hline
2PC        & $1.0\times$ & $2.0\times$ & $2.9\times$ & $4.3\times$ & $5.2\times$ & $6.4\times$ \\ \hline
3PC        & $1.0\times$ & $2.0\times$ & $2.9\times$ & $4.3\times$ & $5.2\times$ & $6.5\times$ \\ \hline
2PC-Dealer & $1.0\times$ & $2.0\times$ & $2.9\times$ & $4.3\times$ & $5.2\times$ & $6.4\times$ \\ \hline
\multicolumn{7}{|c|}{Decoding Time} \\ \hline
2PC        & $1.0\times$ & $2.0\times$ & $2.8\times$ & $4.1\times$ & $4.9\times$ & $5.9\times$ \\ \hline
3PC        & $1.0\times$ & $2.0\times$ & $2.8\times$ & $4.0\times$ & $4.7\times$ & $5.7\times$ \\ \hline
2PC-Dealer & $1.0\times$ & $2.0\times$ & $2.8\times$ & $4.0\times$ & $4.7\times$ & $5.7\times$ \\ \hline
\multicolumn{7}{|c|}{Decoding Comm} \\ \hline
2PC        & $1.0\times$ & $2.0\times$ & $2.8\times$ & $4.3\times$ & $5.1\times$ & $6.1\times$ \\ \hline
3PC        & $1.0\times$ & $2.0\times$ & $2.8\times$ & $4.3\times$ & $4.9\times$ & $6.4\times$ \\ \hline
2PC-Dealer & $1.0\times$ & $2.0\times$ & $2.8\times$ & $4.1\times$ & $4.8\times$ & $5.8\times$ \\ \hline
\end{tabular}
\end{table}

\section{Detailed Hyperparameters for Experiments}
\label{app:hyperparameters}
We performed a best-effort hyperparameter optimization under our compute budget by varying the number of training epochs and batch sizes while keeping the other hyperparamters the same across experiments for a given benchmark.
\tableref{hyperparameters} reports the best configuration we found for each experiment.
We use the same configuration for the ablations, i.e., layer-pruning uses the same hyperparameters as layer-freezing, and head-pruning uses the same parameters as head-merging. Experiments combining 2ReLU with \toolname~(\tableref{marill-prior-approx}) use the same parameters as the corresponding \toolname experiments without 2ReLU.
\begin{table}[t]
    \caption{Batch size and number of epochs for all experiments.} \label{tab:hyperparameters}
    \centering
    \begin{tabular}{lrrrrrr}
    \toprule
    & \multicolumn{2}{c}{MTBench} & \multicolumn{2}{c}{HumanEval} & \multicolumn{2}{c}{WMT22} \\
    & epochs & bsz & epochs & bsz & epochs & bsz \\
    \midrule 
    Fine-tuned & $3$ & $64$ & $2$ & $128$ & $1.5$ & $128$ \\
    \midrule
    LF & $5$ & $64$ & $4$ & $64$ & $1.5$ & $128$ \\
    LoRA/LF+LoRA & $5$ & $8$ & $4$ & $8$ & $1.5$ & $128$ \\
    HM/LF+HM - Stage 1 & $3$ & $8$ & $2$ & $64$ & $1.5$ & $128$ \\
    HM/LF+HM - Stage 2 & $5$ & $64$ & $2$ & $64$ & $1$ & $128$ \\
    \bottomrule
    \end{tabular}
    \label{tab:hyper_paramters}
\end{table}

\section{Limitations} \label{app:limitation}
\textbf{Availability of open-source pre-trained model.}
In this work, we introduce a novel paradigm that shows how the publicly available weights of an open-source pre-trained model can be leveraged to accelerate secure inference.
This makes sense in many settings because the provider doesn't have to go through a very expensive pre-training process, and the best open-source models are among the best models out there~\cite{chiang2024chatbot,liu2024your,berkeley-function-calling-leaderboard,wang2023decodingtrust}.
However, there could be domains that require specialized knowledge which does not benefit from the pre-trained knowledge of the available open-source models.
In such cases, the provider has to pre-train their own model, and layer-freezing and LoRA improvements will no longer apply.
We note that if there is significant relevant public data available for that domain, the provider also has the option to open-source its own pre-trained model to leverage our techniques.

\textbf{Delegation setting.}
In this work, we focus on the secure inference threat models considered by prior work. These works assume that client is one of the MPC participants, and thus, having it evaluate a part of the network locally with layer-freezing actually reduces its overhead. This is because the MPC overhead on each participant is orders higher than plaintext inference~\cite{sigma,mpcformer}.
However, one could also imagine a \emph{weaker threat model} for all of these settings where the client does not participate in the MPC at all. Rather, an additional server is introduced to the MPC with the \emph{additional trust assumption} that it will not collude with the other servers involved in the MPC.
In this case, our layer freezing technique is indeed adding additional overhead on the client, which might not be acceptable in some cases.

\section{Social Impact}
\label{sec:social}
This paper presents work that enables privacy-preserving inference, where both the user's input as well as the service provider's model weights stay private.
While user privacy is needed in many applications and desirable in general, there is a potential concern of model misuse through malicious user prompts.
This is not a fundamental issue though, as the checks that the services perform today on user prompts can also be performed within MPC without revealing them to the service provider. Alternatively, at the cost of additional client overhead, the client could be asked to create a zero-knowledge proof~\cite{zkp} proving that its input satisfies some criteria.

\end{document}